\documentclass[
    ,final            
  ]
  {aipproc}

\layoutstyle{6x9}

\begin{document}

\title{The NMSSM implementation in WHIZARD}


\author{J\"urgen Reuter}{
  address={University of Freiburg, Institute of Physics,
  Hermann-Herder-Str. 3, 79104 Freiburg, Germany}
}

\author{Felix Braam}{
  address={University of Freiburg, Institute of Physics,
  Hermann-Herder-Str. 3, 79104 Freiburg, Germany}
}

\begin{abstract}
The Next-To-Minimal-Supersymmetric extension of the Standard Model
(NMSSM) has been in the focus of extensive studies in the past two
decades. In anticipation of the LHC era, the interest in automatized
tools that can calculate collider signatures has grown. We present the
implementation of the NMSSM into the event generator 
WHIZARD. In addition to a brief review of the implementation, we
discuss the testing and validation procedure. Phenomenological studies
will not be presented here.
\end{abstract}

\maketitle


\section{The Multi-Purpose Event Generator WHIZARD}

WHIZARD~\cite{Kilian:2007gr} is a multi-purpose Monte-Carlo event
generator for the Standard Model (SM) and beyond (BSM). It uses a
multi-channel adaptive  phase space integration based on the VAMP
algorithm~\cite{Ohl:1998jn}, which is called by a highly efficient
phase space grid decomposition relying on heuristics of the underlying
resonance structure of field theoretic scattering amplitudes. The
matrix element are delivered by the generator
O'Mega~\cite{Moretti:2001zz} which avoids in an 
optimal way all redundancies of tree-level scattering amplitudes and
combines gauge-invariant substructures. WHIZARD was the first
generator for full matrix elements for the MSSM~\cite{whiz_susy} and
has been validated by many groups and projects both for supersymetric
models and alternative BSM models~\cite{whiz_bsm}. Among the available
models are half a dozen variants of Little Higgs models,
extra-dimensional models, several (extended) supersymmetric models,
general anomalous gauge, top and Higgs couplings, general
unitarization models for electroweak scattering processes,
noncommutative versions of the Standard Model etc.  

WHIZARD can be accessed via the Hepforge web site:
\url{http://projects.hepforge.org/whizard}. The alpha version of
release 2.0.0 has been finished just before the SUSY conference,
including the first implementation of the Next-to-Minimal Standard
Model (NMSSM) in a multi-particle event generator, which is the main
topic here. We briefly want to review the new features of version
2.0.0, which brings a major improvement over the older version 1. 
WHIZARD has been completely restructured in a fully object-oriented
way. It allows for an arbitrary setup of structure functions,
comprises a fully flexible script language for defining scales, cuts,
analyses variables as arbitrary functions of kinematic variables and
particle lists, it contains a rudimentary own parton shower (which
will be further developed during the development of WHIZARD 2 and
connected with a module for simulating multiple interactions which is
right now being developed). WHIZARD is able to write out standard
event formats like LHA, LHEF, STDHEP, HepMC etc. It comes with its own
graphics analysis tool, while an interface to ROOT will also be available
in WHIZARD 2. Also, an interface to DELPHES~\cite{Ovyn:2009tx} for a
fast detector simulation is under way.   

Although, WHIZARD is able to deliver matrix elements for $2\to 16$
processes (or higher multiplicities) and integrate processes with at
least up to ten particles in the final states, this might not be
sufficient for BSM models with a discrete parity symmetry as suggested
by the existence of dark matter. Hence, WHIZARD 2 is also able to
integrate and simulate long cascade decay chains with full spin
correlations. A further improvement is a full-fledged interface to
FeynRules~\cite{feynrules} for a more automatized inclusion of new
physics models for both WHIZARD versions 1 and 2. 


\section{The NMSSM}

The Next-to-Minimal Supersymmetric Standard Model (NMSSM) is a
viable extension of the MSSM in the Higgs sector of the
model~(see e.g. the references in~\cite{Accomando:2006ga}). The main
motivation for the NMSSM is a possible solution to the so-called $\mu$
problem, the fact, that the supersymmetric parameter $\mu$ in the
superpotential is dimensionful and should in principle be of the order
of the SUSY breaking scale. A working electroweak symmetry breaking
demands this parameter, however, in the ball park of a few hundred
GeV. The NMSSM tackles this problem by enlarging the particle
spectrum by a single left-chiral superfield $S$, being a singlet under
the SM gauge group. Their are two general options to generate a
quartic potential term for the scalar part of $S$: either a $D$ term
of an additional $U(1)$ guage symmetry (like  e.g. in the
PSSSM~\cite{Kilian:2006hh}) or by a $F$ term from a cubic
superpotential term (conventional NMSSM). For the conventions, we
follow those of the SUSY Les Houches Accord 2~\cite{slhaetc} (for a
recent  review on NMSSM conventions
cf. also~\cite{Accomando:2006ga}). This is a quite 
general approach, but neglecting possible CP, $R$-parity, or flavor
violation. The suporpotential of the NMSSM is given by 
\begin{equation}
  \label{eq:nmssmsup}
  W_{NMSSM} = W_{MSSM} - \epsilon_{ab}\lambda {S} {H}^a_1 {H}^b_2 +
  \frac{1}{3} 
\kappa {S}^3 + \mu' S^2 +\xi_F S \ , 
\end{equation}
where we neglect an explicit $\mu'$ term and the Fayet-Iliopoulos term
for the singlet superfield. The most general soft-breaking terms for
the NMSSM  
\begin{equation}
 \label{eq:nmssmsoft}
 V_\mathrm{soft} = V_{2,MSSM} + V_{3,MSSM} + m_\mathrm{S}^2 | S |^2 +
 (-\epsilon_{ab}\lambda A_\lambda {S} {H}^a_1 {H}^b_2 +
 \frac{1}{3} \kappa A_\kappa {S}^3
 + m_{S}'^2 S^2 +\xi_S S
 + \mathrm{h.c.}) \ , 
\end{equation}
using the conventions from~\cite{slhaetc}. 

The field content of the NMSSM is almost the same as for the MSSM,
except for an additional scalar and pseudoscalar Higgs boson, denoted
by $H_3^0$ and $A_2^0$, as well as a fifth neutralino,
$\tilde{\chi}_5^0$ coming from the additional singlino component.


\section{Implementation and Validation}

As mentioned in the section about the model before, we are sticking to
the SLHA 2 conventions. This fixes the uncertainties in signs and
phases. As for the MSSM, the WHIZARD implementation uses explicitly
positive masses for charginos and neutralinos, and puts signs and
complex phases instead in the corresponding mixing matrices. In
contrast to the MSSM, the NMSSM implementation is more flexible than
the MSSM one (although now thanks to Bj\"orn Herrmann there is a
completely general MSSM implementation in WHIZARD), as it allows not
only for a full CKM matrix but also for a left-/right mixing for all
generations, but no inter-generational mixing. The generalization to
the CP-non-conserving case is easily possible. 

The NMSSM is a whole is an incredibly complicated model with order
6,700 couplings (including quartic and Goldstone couplings, most of
each are fortunately not of phenomenological importance). It is
mandatory to check an implementation as far as possible. For this
task, we followed the strategy given in~\cite{Hagiwara:2005wg}:
unitarity checks for $2\to 2$ and $2\to 3$ scattering amplitudes have
been performed, and we tested Ward- and Slavnov-Taylor identities for
both gauge and supersymmetry. Another check was to reproduce the
correct MSSM limit for the NMSSM, namely letting the trilinear singlet
Higgs coupling and the cubic singlet coupling approach zero, sending
the singlet vev to infinity while keeping the combination of the
latter two fixed to $\mu$: $\lambda \to 0$, $\kappa \to 0$, $\left< S
\right> \to  \infty$, $\left< S \right> \lambda \to \mu$.  

\begin{figure}
  \includegraphics[height=.4\textheight]{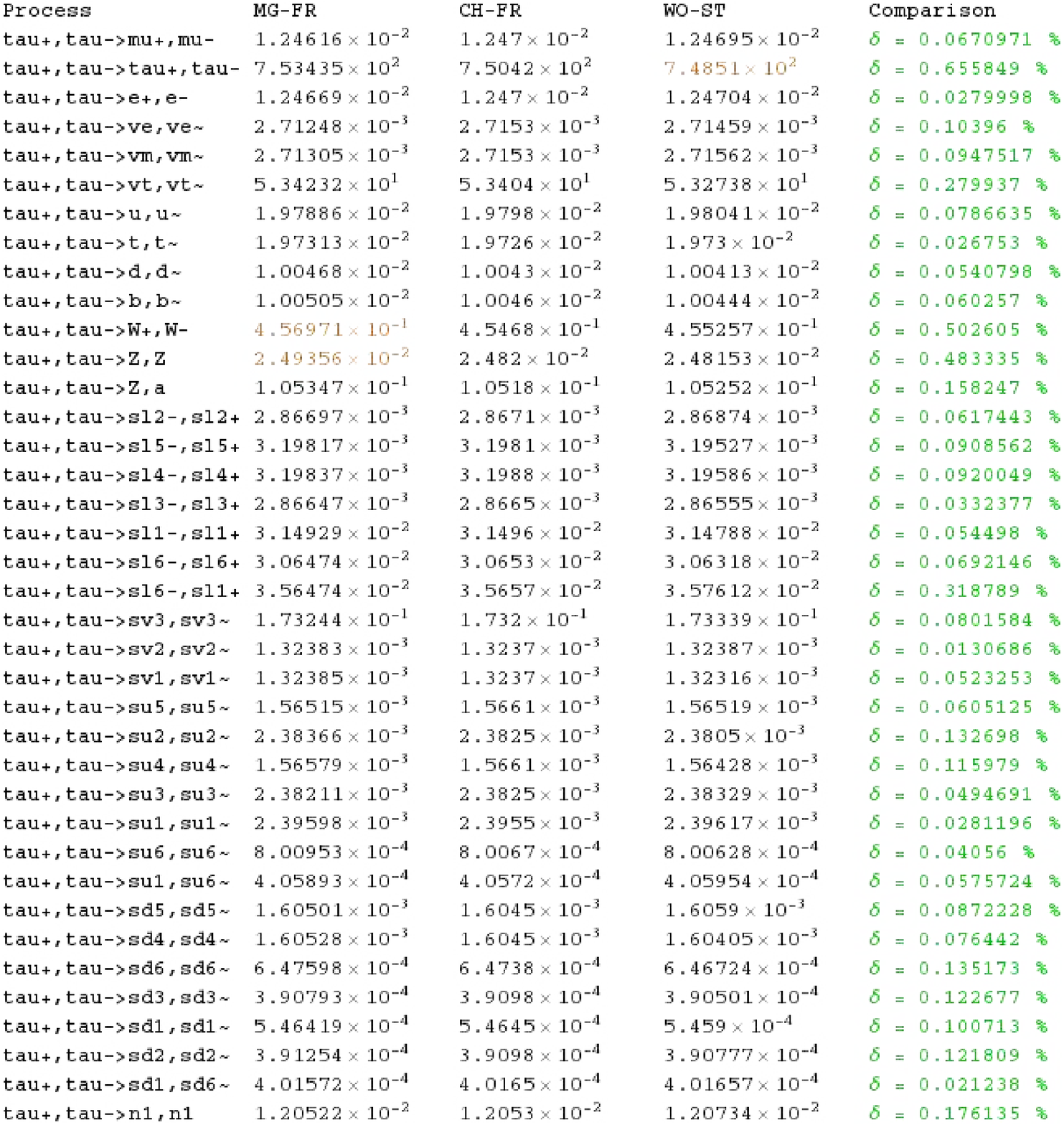}
  \hspace{5mm}
  \includegraphics[height=.4\textheight]{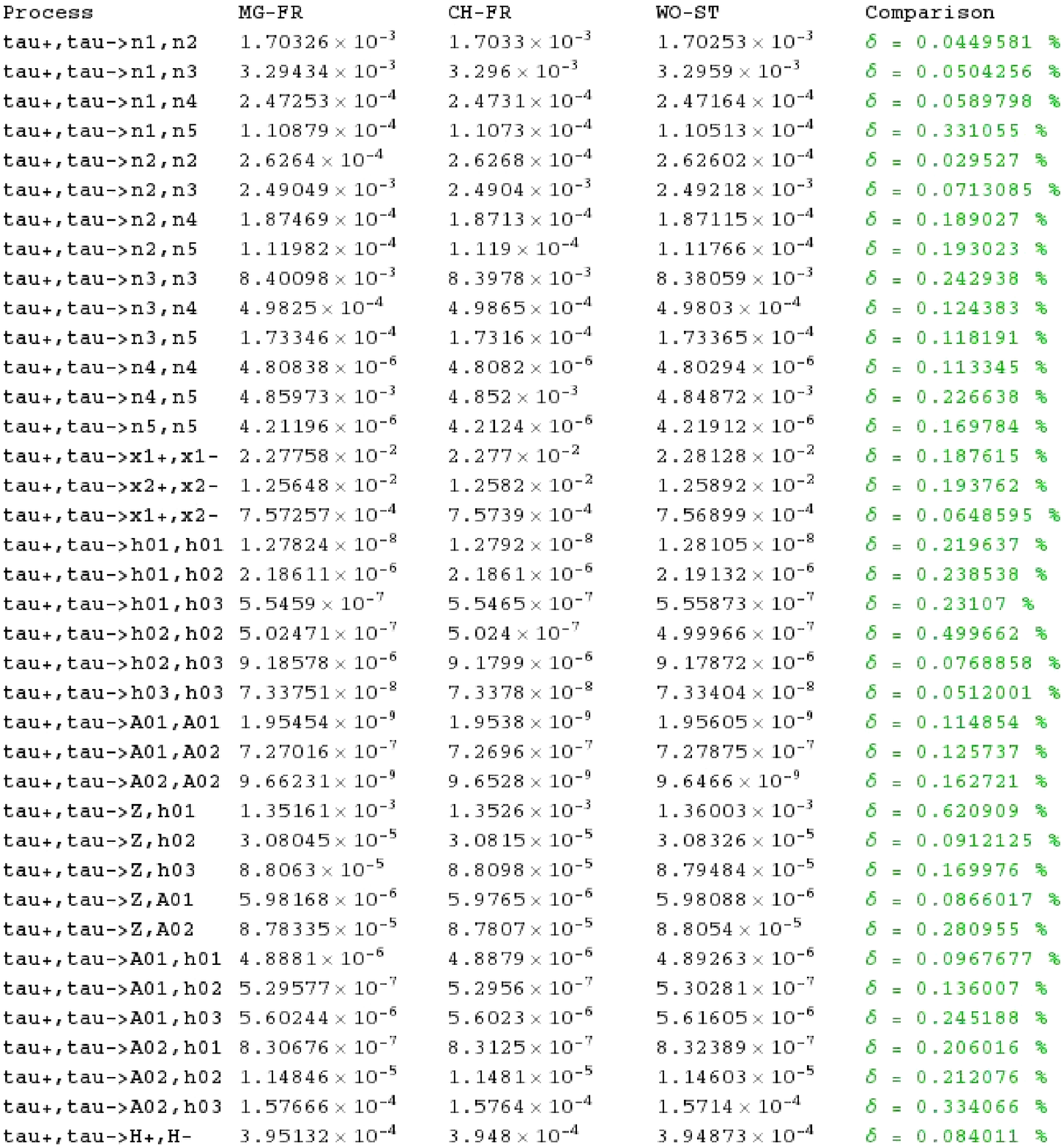}
  \caption{\label{tab1} Comparison between WHIZARD and the FeynRules 
  implementation within Madgraph. The green color shows
  agreement, at least up to the Monte Carlo integration error.
  We used $\sqrt{s} = 3$ TeV to be above all thresholds.}
\end{figure}

What is by far one of the most stringent tests is the comparison with
an independent implementation. For that purpose we are using a
FeynRules generated NMSSM model file for MadEvent/Madgraph~\cite{mgme}
as well as CalcHEP~\cite{Pukhov:2004ca}. Table~\ref{tab1} shows the
comparison for $\tau^+\tau^-$ initial state processes as an
example. The corresponding SLHA 2 input file is left out here due to
reasons of limited space. Our comparison contains far more than 600
processes, showing full agreement between the two implementations.


\section{Conclusions}

We presented the complete implementation of the NMSSM implementation
into the multi-purpose event generator WHIZARD. A vast sample of
consistency checks and comparisons with independent implementations
show the correctness of the code. WHIZARD is available as a modern
high-level tool for NMSSM signal and background simulations for the
LHC era. A detailed version of the comparison and validation of the
NMSSM implementations will appear soon~\cite{bfr}.


\begin{theacknowledgments}
The authors are supported by the Ministerium f\"ur Bildung und Kultur
of the state Baden-W\"urttemberg by the program ZO IV and the German
Research Society (DFG) under grant no. Re 2850/1-1. JR would like to
thank to Aspen Center for Physics for their hospitality. We are also
grateful to Ulrich Ellwanger for valuable remarks and discussions, as
well as Benjamin Fuks, Neil Christensen, Claude Duhr and Christian
Speckner for the FeynRules collaboration.
\end{theacknowledgments}



\bibliographystyle{aipproc}   

\bibliography{sample}



\end{document}